\documentclass[twoside]{article}
\usepackage{epsfig.sty}
\newcommand{\beq}{\begin{eqnarray}}
\newcommand{\eeq}{\end{eqnarray}}
\begin{document}
\title{Heavy quark state production and suppression via Xe-Xe 
collisions at $\sqrt{s_{pp}}$=5.44 TeV}
\author{Leonard S. Kisslinger$^{1}$\\
Department of Physics, Carnegie Mellon University, Pittsburgh PA 15213 USA.\\
Debasish Das$^{2}$\\
Saha Institute of Nuclear Physics, A CI of Homi Bhabha National Institute,\\
 1/AF, Bidhan Nagar, Kolkata 700064, INDIA.}
\date{}
\maketitle

\vspace{-1cm}

\noindent
1) kissling$@$andrew.cmu.edu \hspace{1cm} 2) debasish.das@saha.ac.in; dev.deba@gmail.com

\begin{abstract}
   We estimate the differential rapidity cross sections for
$J/\Psi$, $\Psi(2S)$, $\Upsilon(1S)$, $\Upsilon(2S)$, and $\Upsilon(3S)$ 
production and $\Psi(2S)$ to $J/\Psi(1S)$, $\Upsilon(3S)$ to $\Upsilon(1S)$ 
suppression via Xe-Xe collisions at proton-proton energy $\equiv \sqrt{s_{pp}}$
= 5.44 TeV. For the $\Psi(2S)$, $\Upsilon(3S)$  states we use the mixed 
heavy quark hybrid theory, with these states being approximately 50\% standard
and 50\% hybrid charmonium and bottomonium meson states.
\end{abstract}
\noindent
PACS Indices:12.38.Aw,13.60.Le,14.40.Lb,14.40Nd
\vspace{1mm}


\section{Introduction}

This new work on Xe-Xe collisions is based on the heavy quark state 
production formalism in Pb-Pb collisions at $\sqrt{s_{pp}}$=p-p energy=5.02
TeV\cite{kd17}.
\vspace{4mm}

We use the standard model for $J/\Psi$, $\Upsilon(1S)$, and $\Upsilon(2S)$ 
states and the mixed hybrid theory\cite{lsk09} for  $ \Psi(2S), \Upsilon(3S)$ 
states. Here we have also explored $\Psi(2S)$ to $J/\Psi(1S)$ and 
$\Upsilon(3S)$ to $\Upsilon(1S)$ suppression scenario as studied in p-Pb 8 TeV 
collisions\cite{kd16}. Recently the LHCb collaboration\cite{LHCb18} found a 
suppression of about 40\% for $\Upsilon(nS)$ states produced in p-Pb collisions.
\vspace{4mm} 

In section 2 heavy quark hybrid states and mixed heavy quark hybrid states are
reviewed. Both the bottom quark $b$ and the charm quark $c$ are 
needed, with masses\cite{ppb} $m_c \simeq$ 1.27 GeV and$m_b \simeq$ 4.18 GeV. 
Also the anti-quarks $\bar{c}$ and  $\bar{b}$ are needed.
As discussed in section 2  the state $J/\Psi(1S)$ is $ |c\bar{c}(1S)>$ while
state $\Psi(2S)$ is a mixed hybrid $c$ meson. The state $\Upsilon(3S)$ is
a mixed mixed hybrid $b$ meson.
\vspace{4mm}

In section 3, $\Psi$ and $\Upsilon$ production in Xe-Xe collisions, our new
work on heavy quark state production is based on the methods used
in  heavy quark state production in Cu-Cu and Au-Au collisions at
$\sqrt{s_{pp}}$=200 GeV\cite{klm14} which used the color octet 
model\cite{cl96,bc96,fl96}. Prior to the article\cite{kd16} $\Psi(2S)$ and
$\Upsilon(3S)$ suppression in p-Pb collisions with E=$\sqrt{s_{pp}}$= 5.02 TeV 
was estimated\cite{lsk16} and reviewed\cite{lskdd16}. Also, the ALICE
collaboration studied $J/\Psi$ production via Xe-Xe collisions at
$\sqrt{s_{NN}}$= 5.44 TeV\cite{alice18}.
\vspace{4mm}

In section 4 $\Psi$ and $\Upsilon$ suppression in Xe-Xe collisions and mixed
hybrid theory the suppression, $S_A$, of charmonium or bottomonium states 
was  estimated for Pb-Pb and  p-Pb collisions.
\vspace{4mm}

In subsection 4.1 $\Psi(2S)$ suppression in Xe-Xe collisions is reviewed.
\vspace{4mm}

In subsection 4.2 $\Upsilon(3S)$ suppression in Xe-Xe collisions is estimated
to be moderate.
\newpage
\section{Mixed heavy quark hybrid states}
\vspace{4mm}

The starting point of the method of QCD sum rules\cite{sz79} is the correlator
\beq
\label{cor}
       \Pi^A(x) &=&  \langle | T[J_A(x) J_A(0)]|\rangle \; ,
\eeq
with $| \rangle$ the vacuum state and the current $J_A(x)$ creates the 
states with quantum numbers A.
\vspace{4mm}

With $c$, $\bar{c}$ and $g$ a charm quark, an anti-charm quark and a 
gluon, for the charmonium states, $J_{c}$ is
\beq
\label{11}
        J_{c} &=& f J_{c\bar{c}} + \sqrt{1-f^2} J_{c \bar{c} g} \; , 
\eeq
where $J_{c \bar{c}}$ creates a normal charmonium state and $J_{c \bar{c} g}$ 
creates a hybrid state with an active gluon.
\vspace{4mm}

The charm quark $c$ needed, with mass\cite{ppb} $m_c \simeq$ 1.27 GeV
\vspace{4mm}

Using QCD sum rules it was shown that $f \simeq -\sqrt{2}$ for the 
$\Psi(2S)$ and $\Upsilon(3S)$ heavy quark meson states amd $f \simeq 1.0$ for 
the other charmonium and bottomonium states\cite{lsk09}.
\vspace{4mm}

{\bf Therefore,} 
\beq
\label{psi}
     |J/\Psi(1S)> &\simeq& |c\bar{c}(1S)>  \\
    |\Psi(2S)> &\simeq& -\sqrt{2} |c\bar{c}(1S)>+\sqrt{2}|c\bar{c}g(2S)> 
\nonumber \; ,
\eeq
with the $\Psi(2S)$ being a mixed hybrid meson.
\vspace{4mm}

Similarly, it was shown with $b$ and $\bar{b}$ a bottom and anti-bottom
quark, with masse\cite{ppb} $m_b \simeq$ 4.18 GeV, that

\beq
\label{up}
  |\Upsilon(3S)> &\simeq& -\sqrt{2} |b\bar{b}(3S)>+\sqrt{2}|b\bar{b}g(3S)> \; ,
\eeq
\vspace{4mm}

After a larger production of $\Psi(2S)$ states via high energy collisions 
than standard model predictions\cite{cdf},  and the
anomalous production of sigmas in the decay of $\Upsilon(3S)$ to  
$\Upsilon(1S)$ was found\cite{vogel}, the solution of these anomalies was found
in the mixed hybrid theory\cite{lsk09}.  The $\Psi(2S)$ state was found to be
\beq
\label{1}
        |\Psi(2S)>&=& \alpha |c\bar{c}(2S)>+\sqrt{1-\alpha^2}|c\bar{c}g(2S)> 
\; ,
\eeq
where $c$ is a charm quark, and the $\Upsilon(3S)$ state was found to have 
the form
\beq
\label{2}
    |\Upsilon(3S)>&=& \alpha |b\bar{b}(3S)>+\sqrt{1-\alpha^2}|b\bar{b}g(3S)> 
\; ,
\eeq
where $b$ is a bottom quark and $\alpha = -0.7 \pm 0.1$, as in Ref\cite{sz79}.
\vspace{4mm}

That is, these states have approximately a 50\% probability of being a standard 
quark-antiquark, $|q\bar{q}>$, meson, and a 50\% probability of a hybrid, with 
the $|q\bar{q}g>$ a color octet and an active gluon, which we use for both the 
production and suppression of heavy quark states via Xe-Xe collisions.
\vspace{4mm}

It is important to use the mixed hybrid theory as the $\Psi(2S),\Upsilon(3S)$ 
cross sections are enhanced by a factor of $\pi^2/4$\cite{klm11}.
\vspace{4mm} 

This is in the following section on $\Psi$ and $\Upsilon$ production in Xe-Xe
collisions.
\newpage
\section{$\Psi$ and $\Upsilon$ production in Xe-Xe
  collisions with $\sqrt{s_{pp}}$ = 5.44 TeV}

 The differential rapidity cross section for the production of a heavy
quark state $\Phi$ with helicity $\lambda=0$ (for unpolarized  
collisions\cite{klm11}) in the color octet model in Xe-Xe collisions is given 
by\cite{klm14}
\beq
\label{3}
   \frac{d \sigma_{AA\rightarrow \Phi(\lambda=0)}}{dy} &=& 
   R^E_{AA} N^{AA}_{bin}< \frac{d \sigma_{pp\rightarrow \Phi(\lambda=0)}}{dy}>
\; ,
\eeq
where $R^E_{AA}$ is the product of the nuclear modification factor $R_{AA}$
and $S_{\Phi}$, the dissociation factor after the state $\Phi$ is formed 
(see FIG.3 in Ref\cite{star09}). $R_{AA}$ is defined in Ref\cite{star02} as
\beq
\label{RAA}
  R_{AA}(p)&=& \frac{d^2  N^{AA}_{bin}/dpd\eta}{T_{AA} d^2 \sigma^{NN}/dpd\eta}
\; ,
\eeq
where $N^{AA}_{bin}$ is the number of binary collisions in the AA collision,
$\eta$ is the pseudo-rapidity\cite{star01}, $T_{AA}= N^{AA}_{bin}/\sigma^{NN}$, 
where $\sigma^{NN}$ is the cross section for N-N collisions and
 $< \frac{d \sigma_{pp\rightarrow \Phi(\lambda=0)}}{dy}>$ is the 
differential rapidity cross section for $\Phi$ production via nucleon-nucleon 
collisions in the nuclear medium. For Xe-Xe collisions we use  $R^E_{XeXe}$ =0.5
as $R^E_{AA}\simeq 0.5$ both for Cu-Cu\cite{star09,phenix08} and 
Au-Au\cite{phenix07,star07,kks06} and $R^E_{PbPb}$ =0.5 was used in 
Ref\cite{kd17}. $N^{XeXe}_{bin}$ is the number of binary collisions in Xe-Xe
collisions, and $< \frac{d \sigma_{pp\rightarrow \Phi(\lambda=0)}}{dy}>$ is the 
differential rapidity cross section for $\Phi$ production via nucleon-nucleon 
collisions in the nuclear medium. The number of binary collisions for Pb-Pb
from Ref\cite{sbstar07} used in Ref\cite{kd17} was $N^{PbPb}_{bin}\simeq$  260. 
\vspace{4mm}

The cross sections at $\sqrt{s_{pp}}\simeq$5.0 TeV\cite{lke17} $\sigma(b)\simeq$
 7.66, 3.34, 5.61 for PbPb, CuCu, XeXe. 
Using $N^{PbPb}_{bin}$= 260, for Cu-Cu collisions 
$N^{CuCu}_{bin}\simeq$  51.5\cite{klm14,sbstar07}. From  $N^{PbPb}_{bin}$= 260 and
the Xe-Xe, Pb-Pb cross sections we estimate $N^{XeXe}_{bin}\simeq$ 216. Therefore
in Eq(\ref{2}) $R^E_{AA} N^{AA}_{bin} \rightarrow R^E_{XeXe} N^{XeXe}_{bin}\simeq 108$.

The differential rapidity cross section for pp collisions 
in terms of $f_g$\cite{klm11}, the gluon distribution function is
\beq
\label{4}
     < \frac{d \sigma_{pp\rightarrow \Phi(\lambda=0)}}{dy}> &=& 
     A_\Phi \frac{1}{x(y)} f_g(\bar{x}(y),2m)f_g(a/\bar{x}(y),2m) 
\frac{dx}{dy} \; ,
\eeq  
where from Ref\cite{klm11}  $A_\Phi=1.26 \times 10^{-6}$ nb for 
$\Phi$=$J/\Psi$ and $3.4 \times 10^{-8}$ nb for $\Phi$=$\Upsilon(1S)$; and
$a=4 m^2/s =3.6 \times 10^{-7}$ for charmonium and $4.0 \times 10^{-6}$ for 
bottomonium.
\vspace{4mm}

 The function $\bar{x}$, the effective parton x in a nucleus (A), is given in  
Refs\cite{vitov06},\cite{sharma09}:
\beq
\label{barx}
         \bar{x}(y)&=& (1+\frac{\xi_g^2(A^{1/3}-1)}{Q^2})x(y) \nonumber \\
   x(y) &=& 0.5 \left[\frac{m}{\sqrt{s_{pp}}}(\exp{y}-\exp{(-y)})+
\sqrt{(\frac{m}{\sqrt{s_{pp}}}(\exp{y}-\exp{(-y)}))^2 +4a}\right] \;,
\eeq
with\cite{qiu04} $\xi_g^2=.12 GeV^2$ and $Q^2$ defined in Ref\cite{vitov06}.  
\vspace{4mm}

Therefore for Xe with $A\simeq 132$,
$Q^2\simeq 10.18 \; GeV^2$
\beq
\label{barx2}
         \bar{x}(y)&=& 1.048 x(x) \; .
\eeq

For $\sqrt{s_{pp}}$=5.44 TeV the gluon distribution function $f_g$\cite{CTEQ6,klm11} 
is
\beq
\label{fg} 
    f_g(\bar{x}(y),2m)&=& 1334.21-67056.5 \bar{x}(y)+887962.0 (\bar{x}(y))^2
\; .
\eeq

With $\Psi(2S),\Upsilon(3S)$ enhanced by $\pi^2/4$\cite{klm11} the 
differential rapidity cross sections are shown in the following figures.
The absolute magnitudes are uncertain, and the shapes and relative magnitudes 
are our main predictions. In Eq(\ref{barx}) $m=m_c=$ 1.5 GeV for charmonium
and $m=m_b=$ 5 GeV for bottomonium quarks.

\newpage

\begin{figure}[ht]
\begin{center}
\epsfig{file=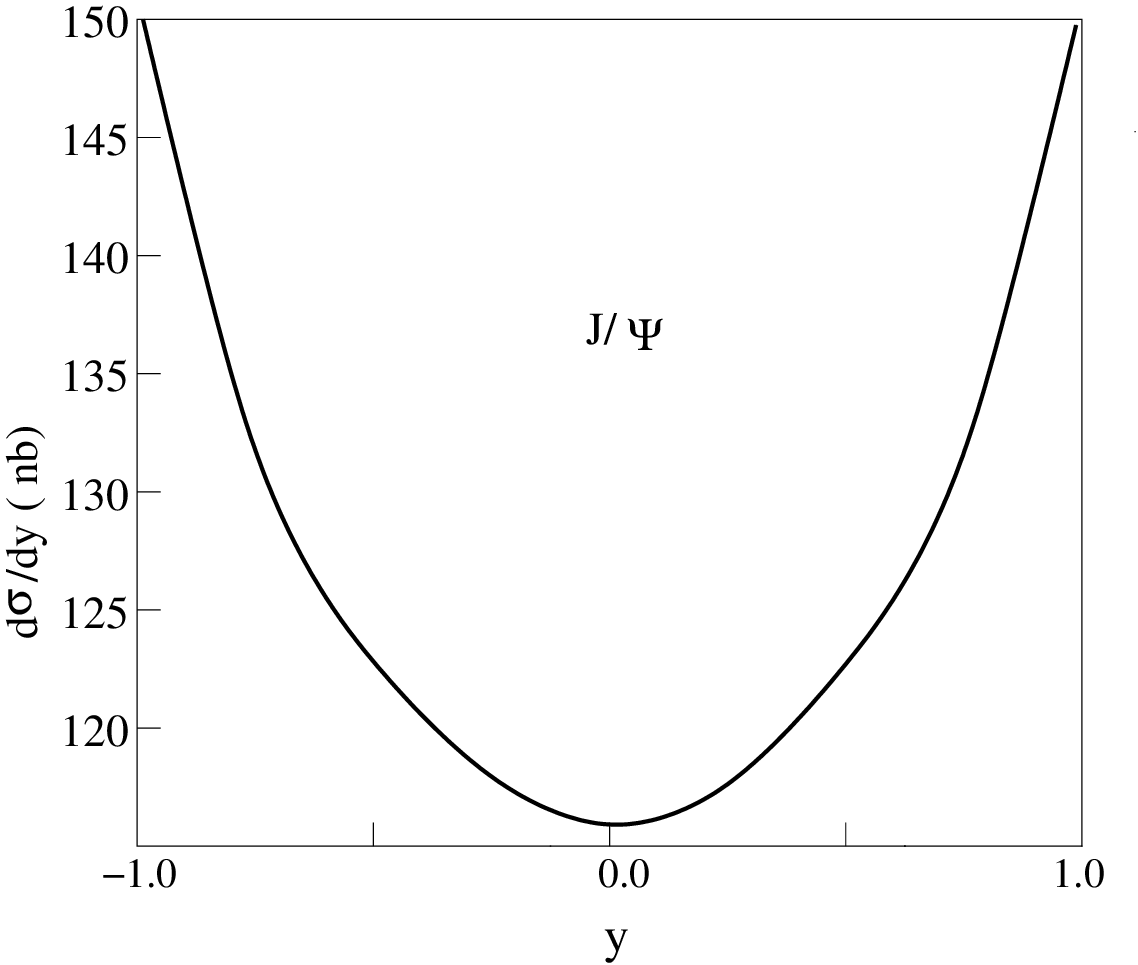,height=7 cm,width=10cm}
\caption{d$\sigma$/dy for 2$m_c$=3 GeV, $\sqrt{s_{pp}}$=5.44 TeV Xe-Xe 
collisions producing $J/\Psi$ with $\lambda=0$}
\label{Figure 1}
\end{center}
\end{figure}
\vspace{1cm}

\begin{figure}[ht]
\begin{center}
\epsfig{file=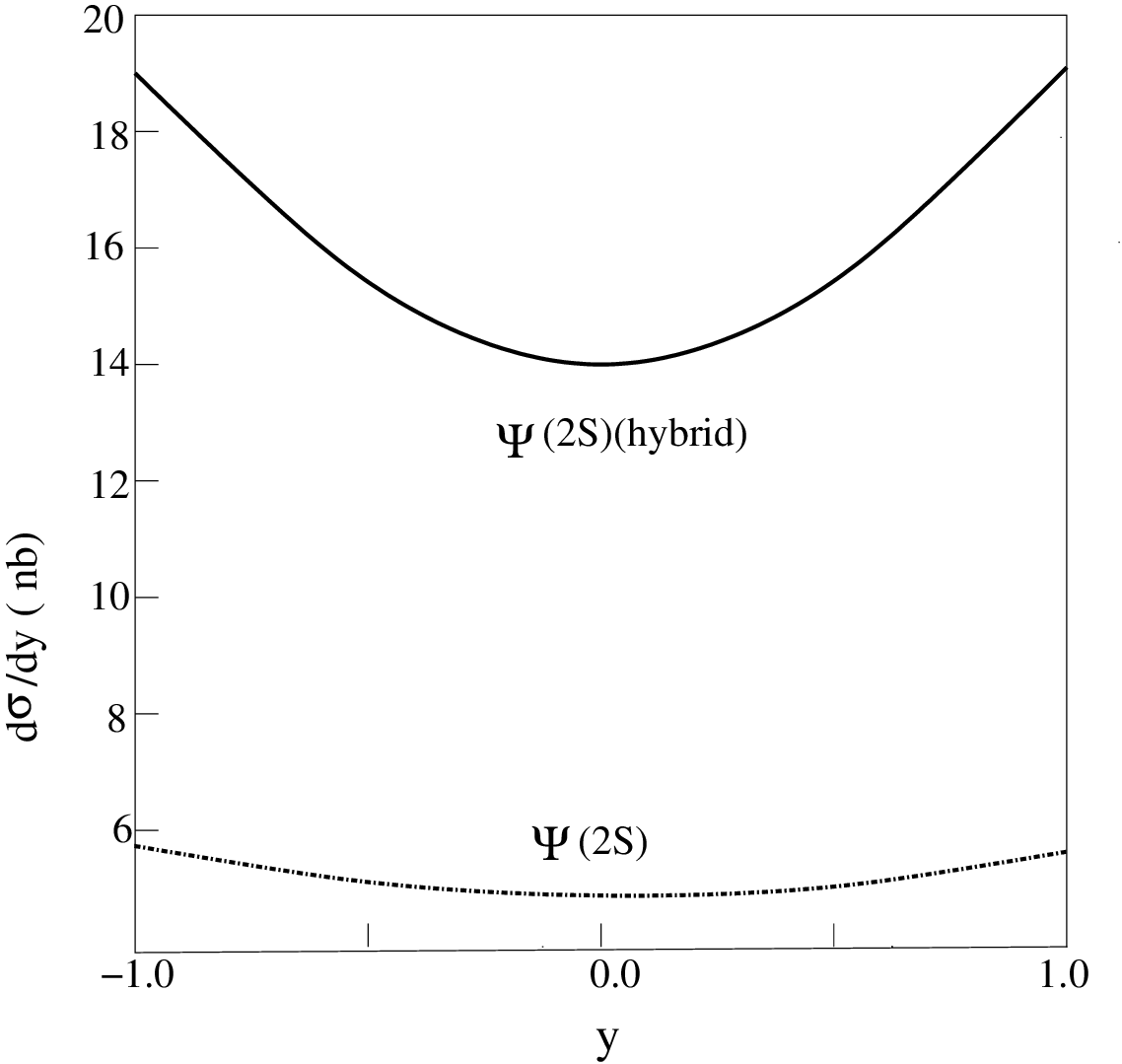,height=7 cm,width=10cm}
\caption{d$\sigma$/dy for 2$m_c$=3 GeV, $\sqrt{s_{pp}}$=5.44 TeV Xe-Xe 
collisions producing $\Psi(2S)$, hybrid theory, with $\lambda=0$. The dashed 
curve is for the standard $c\bar{c}$ model.}
\label{Figure 2}
\end{center}
\end{figure}
\newpage

\begin{figure}[ht]
\begin{center}
\epsfig{file=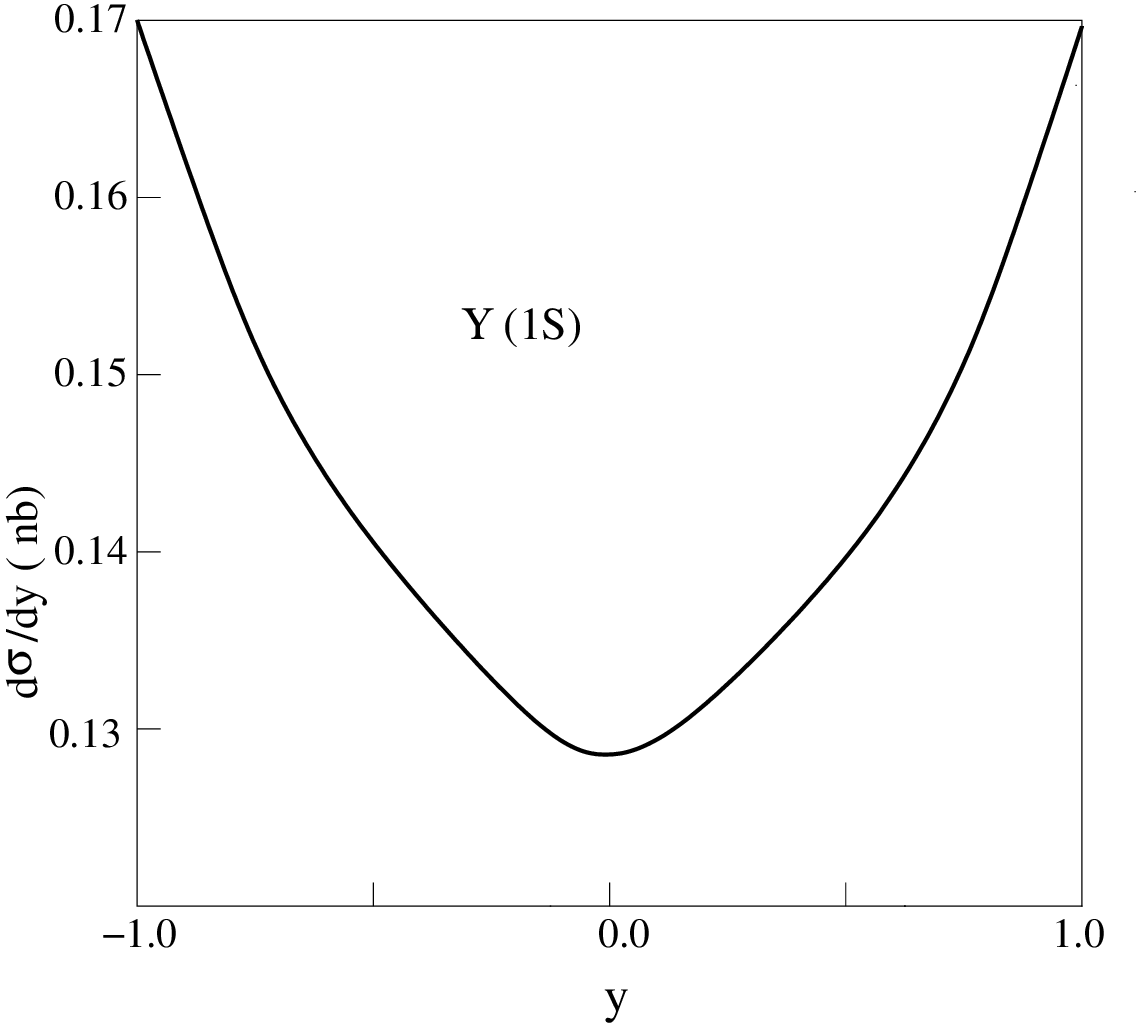,height=7 cm,width=10cm}
\caption{d$\sigma$/dy for 2$m_b$=10 GeV, $\sqrt{s_{pp}}$=5.44 TeV Xe-Xe 
collisions producing $\Upsilon(1S)$, $\lambda=0$}
\label{Figure 3}
\end{center}
\end{figure}
\vspace{1cm}

\begin{figure}[ht]
\begin{center}
\epsfig{file=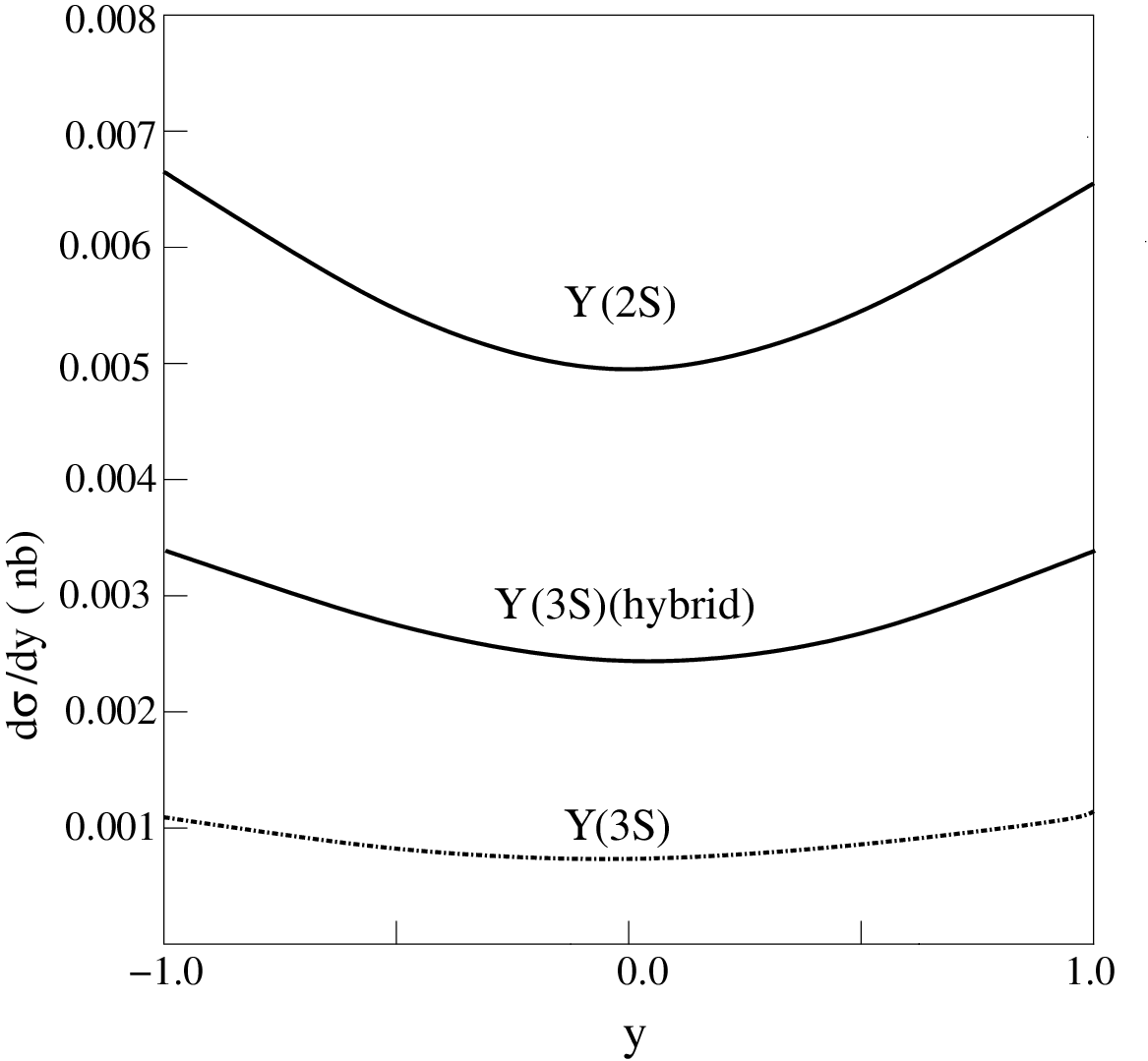, height=7 cm,width=10cm}
\caption{d$\sigma$/dy for 2$m_b$=10 GeV, $\sqrt{s_{pp}}$=5.44 TeV Xe-Xe 
collisions producing with $\lambda=0$ $\Upsilon(2S)$ and $\Upsilon(3S)$(hybrid).
For $\Upsilon(3S)$ the dashed curve is for the standard $b\bar{b}$ model.}
\label{Figure 4 }
\end{center}
\end{figure}

\newpage
\section{$\Psi$ and $\Upsilon$ suppression in Xe-Xe
 collisions and mixed hybrid theory}

The suppression, $S_A$, of charmonium or bottomonium states is given by the 
interaction of the meson with nucleons as it traverses the nucleus.
From Ref\cite{ks96}

\beq
\label{SA}
    S_A &=& e^{-n_o\sigma_{\Phi N} L} \; ,
\eeq
where $\Phi$ is a standard $c\bar{c}$, hybrid $c\bar{c}g$ charmonium meson,
or a standard $b\bar{b}$, $b\bar{b}g$ bottomonium meson.  $L$ is the length 
the path of $\Phi$ in nuclear matter, $n_o=0.17$ fm$^{-3}$ is the nuclear 
matter density, and $\sigma_{\Phi N}$ is the $\Phi-N$ cross section. For
Xe-Xe collisions $L\simeq$ 12 fm. 

  $S_A$ was estimated for Pb-Pb collisions\cite{kd17}  and  p-Pb 
collisions\cite{lsk16}. From Ref\cite{lsk16} $\sigma_{\Phi N}$ for
$\Phi=c\bar{c},c\bar{c}g,b\bar{b},b\bar{b}g$ were estimated to be (with a
mb $\rightarrow$ fm$^2$ correction)
\beq
\label{sigma}
 \sigma_{c \bar{c} N} &\simeq & 3.2 \times 10^{-2} \mathrm{\;fm}^2 \nonumber \\
 \sigma_{c\bar{c}g N} &\simeq & 6.5 ^2 \mathrm{\;fm}^2 \nonumber \\
 \sigma_{b \bar{b} N} &\simeq & 2.9 \times 10^{-3} \mathrm{\;fm}^2 \nonumber \\
 \sigma_{b\bar{b}g N} &\simeq & 0.59 \mathrm{\;fm}^2 \; .
\eeq

Note that the present results for $n_o\sigma_{\Phi N} L$ differ from those in 
Ref\cite{lsk16} for Pb-Pb suppression by a factor of about 1.25 as 
$L\simeq$ 15 fm for Pb and 12 fm for Xe.
\subsection{$\Psi(2S)$ suppression in Xe-Xe collisions}

 From Eq(\ref{sigma}) for $J/\Psi=\Psi(1S)=|c \bar{c};1S>$ 
\beq
\label{Psi1S-nsL}
         n_o\sigma_{c \bar{c} N} L &\simeq& 0.065 \nonumber \\
        S_A^{c\bar{c}} &=& e^{- n_o\sigma_{c \bar{c} N} L} \simeq 1.0 \; ,
\eeq 
so the $J/\Psi=\Psi(1S)$ meson is not suppressed in Xe-Xe collisions.
 For a charmonium hybrid meson $c \bar{c}g$ with L=12 fm
\beq
\label{SAccg}
      n_o\sigma_{c\bar{c}g N} L&\simeq& 13.3 \nonumber \\
   S_A^{c\bar{c}g} &\simeq&e^{-13.3} \simeq 0.0  \; .
\eeq

 Since from Eq(\ref{1}) the $\Psi(2S)$ is 50\% standard and 50\% hybrid
\beq
\label{SAPsi-2}
    S_A^{\Psi(2S)} &\simeq& (1+0.0)/2 \simeq  0.5 \;.
\eeq

  With the definition of $ R^{\Psi(2S)-J/\Psi(1S)}$ the ratio of the suppression of 
the $\Psi(2S)$ to the $\Psi(1S)$ states, from Eqs(\ref{Psi1S-nsL},\ref{SAccg})
\beq
\label{SAratio}
    R^{\Psi(2S)-J/\Psi(1S)}|_{Xe-Xe-theory}&\simeq& 0.5 \; ,
\eeq

The experimental result\cite{PHENIX13} for d-Au collisions and\cite{ALICE14} for p-Pb collisions is
\beq
\label{exp-suppression}
    R^{\Psi(2S)-J/\Psi(1S)}|_{exp}&\simeq& 0.65 \pm 0.1 \,
\eeq
which is somewhat larger than $R^{\Psi(2S)-J/\Psi(1S)}|_{Xe-Xe-theory}$.

\subsection{$\Upsilon(3S)$ suppression in Xe-Xe collisions}

 From Eq(\ref{sigma}) for $\Upsilon(1s)=|b \bar{b};1S>$ 
\beq
\label{Upsilon1S2S-nsL}
         n_o\sigma_{b \bar{b} N} L &\simeq& 0.0 \nonumber \\
        S_A^{b\bar{b}} & \simeq& 1.0. \; ,
\eeq 
so the $\Upsilon(1S)$ and $\Upsilon(2S)$ are not suppressed.

For a hybrid bottomonium meson $b \bar{b}g$ from Eq(\ref{sigma})
\beq
\label{SAbbg}
      n_o\sigma_{\bar{b}g N L}&\simeq& 1.2 \nonumber \\
   S_A^{b\bar{}g} &\simeq& 0.3 \; .
\eeq
 
For the mixed hybrid $\Upsilon(3S)$
\beq
\label{SAU3}
    S_A^{\Upsilon(3S)} &\simeq& (1+.3)/2 \simeq 0.65 \; .
\eeq

Therefore for Xe-Xe collisions the $\Upsilon(3S)$  suppression is moderate
and could be measured in future CERN LHC experiments .

\section{Conclusions}

In section 2 mixed heavy quark hybrid states are reviewed.
\vspace{4mm}

In section 3 we have extended and reviewed our work on heavy quark state
production in Pb-Pb collisions\cite{kd17} we have studied the differential
rapidity cross sections for $J/\Psi, \Psi(2S)$ and $\Upsilon(nS)(n=1,2,3)$
production via Xe-Xe collisions with $\sqrt{s_{pp}}=5.44 TeV$ using
$R^E_{XeXe} N^{XeXe}_{bin} \simeq 108$. This will give guidance for future CERN
LHC experiments\cite{lke17}.
\vspace{4mm}

In section 4  an extension of estimates of suppression for Pb-Pb
collisions\cite{lsk16} and  p-Pb collisions\cite{kd16} was reviewed.
\vspace{4mm}

In subsection 4.1 we have estimated the suppression of $\Psi(2S)$ compared
to $J/\Psi(1S)$ via Xe-Xe collisions.

\vspace{4mm}

In subsection 4.2 we have estimated the suppression of $\Upsilon(nS)(n=1,2,3)$
for Xe-Xe collisions. We used L=12 fm
rather than 15 fm for Pb-Pb collisions for $n_o\sigma_{\Phi N} L$ to estimate 
suppression $ S_A = e^{-n_o\sigma_{\Phi N} L}$, with $\Phi$= $\Psi(1,2)$ and 
$\Upsilon(1,2,3)$. Since $\Psi(2S)$ and$\Upsilon(3S)$ are mixed hybrid mesons,  
$S_A^{\Psi(2S)}\simeq  (S_A^{c\bar{c}}+S_A^{c\bar{c}g})/2$ and
$S_A^{\Upsilon(3S)}\simeq (S_A^{b\bar{b}}+S_A^{b\bar{b}g})/2$. Our results are that
the  $J/\Psi$,  $\Upsilon(1S)$ and $\Upsilon(2S)$ are not suppressed in Xe-Xe
collisions and $S_A^{\Psi(2S)} \simeq  0.5$, $S_A^{\Upsilon(3S)} \simeq 0.65$
for Xe-Xe collisions.
\vspace{4mm}

The experimental results\cite{PHENIX13} for d-Au collisions and\cite{ALICE14}
for p-Pb collisions for the ratio of $\Psi(2S)$ to $J/\Psi(1S)$ are almost
consistent with our present results for Xe-Xe collisions. From our results
$\Psi(2S)$ and $\Upsilon(3S)$ suppression could be measured in future CERN
LHC Xe-Xe collision
experiments\cite{lke17}.
\vspace{5mm}

\Large
{\bf Acknowledgements}
\normalsize
\vspace{5mm}

Author D.D. acknowledges the facilities of Saha Institute of Nuclear Physics, 
Kolkata, India. Author L.S.K. acknowledges support in part by a grant from
the Pittsburgh Foundation.

\end{document}